\documentclass[12pt]{iopart}
\usepackage{bbm} \usepackage{graphicx} \usepackage{iopams}
\begin{document}
\title[Quantum and classical correlations of intense beams 
of light]{Quantum and classical correlations of intense beams of light via
joint photodetection}
\author{Andrea Agliati and Maria Bondani}
\address{INFM, Unit\`a di Como, Italia}
\author{Alessandra Andreoni}
\address{Dipartimento di Fisica e Matematica, Universit\`a degli Studi 
dell'Insubria, Como, Italia}
\author{Giovanni De Cillis and Matteo G. A. Paris
\footnote{{\rm Electronics address:}{\tt matteo.paris@fisica.unimi.it}
\\ $\quad${\rm URL:}{\tt http://qinf.fisica.unimi.it/~paris}}}
\address{Dipartimento di Fisica dell'Universit\`a di Milano, Italia}
\begin{abstract}
We address joint photodetection as a method to discriminate
between the classical correlations of a thermal beam divided by a
beam splitter and the quantum entanglement of a twin-beam
obtained by parametric downconversion. We show that for intense
beams of light the detection of the difference photocurrent may be
used, in principle, in order to reveal entanglement, while the simple
measurement of the correlation coefficient is not sufficient.
We have experimentally measured the correlation coefficient and
the variance of the difference photocurrent on several classical
and quantum states. Results are in good agreement with
theoretical predictions taking into account the extra noise in
the generated fields that is due to the pump-laser fluctuations.
\end{abstract}
\section{Introduction}\label{s:intro}
Entanglement is a crucial resource in quantum information processing,
quantum communication and quantum measurements. Indeed quantum
correlations lead to important novel effects not achievable by using
classically-correlated states, {\em i.e.} states characterized by
correlations that may be established by using local operations and
classical communication. Quantum information has been initially
developed for discrete quantum variables, {\em i.e.  quantum bits},
which can be implemented optically by means of polarization
single-photon states. However, much attention has been recently devoted
to continuous variable (CV) regime and to multiphoton states of light.
Continuous-spectrum quantum variables may be easier to manipulate
compared to quantum bits by means of linear optical circuits and
homodyne detection \cite{pati,nap04,knill}: this is the case of Gaussian
states of light, {\em e.g.} squeezed- and twin-beam. By using CV one may
carry out nonlocality experiments \cite{ban}, quantum teleportation 
\cite{furu} and generation of multimode entanglement \cite{multi}.
The concepts of quantum cloning \cite{clo} and entanglement
purification \cite{geza} have also been extended to CV, and secure
quantum communication protocols have been proposed \cite{gran}.
\par
Ideal features for implementing quantum information experiments are the
availability of bright and stable entanglement sources, based on
degenerate or nondegenerate optical parametric processes, and the
possibility of an effective characterization of entanglement. In the
case of CV Gaussian entanglement quantum correlations may be
discriminated from classical correlations by using homodyne detection.
However, homodyne detection requires an appropriate mode matching of the
signals with a local oscillator at a beam splitter, a task that may be
particularly challenging in the case of pulsed optical fields. On/off 
photodetection may be also used to characterize Gaussian states, but
its use is limited to states with a small number of photons \cite{fiu,cvp}.
\par
For the reasons of above, in this paper we assess the use of intensity
measurements, in particular joint photodetection, as a method to
discriminate classical correlations from entanglement \cite{ray}. A simple
intensity-based measurement, including the measurement of difference
photocurrent, cannot provide a complete characterization of
entanglement. However, we show that for intense beams of light the
detection of the difference photocurrent may be used, in principle, in
order to reveal entanglement, while the simple measurement of the
correlation coefficient is not sufficient. In particular, joint
photodetection can be useful to discriminate the entanglement of
twin-beam from correlations of thermal sources in the mesoscopic regime.
We have experimentally measured the correlation coefficient and the
variance of the difference photocurrent on several classical and quantum
states. Results are in good agreement with theoretical predictions if
one takes into account the extra noise in the generated states caused by
the pump laser fluctuations.
\par
This work may also contribute to the recent debate on the use of classical
and quantum correlations in imaging and on the necessity of entanglement
for extracting the information \cite{qim}. Our results indicate that any
method only based on correlation measurements cannot be improved using
entanglement instead of classical correlations.
\par
The paper is structured as follows. In Section \ref{s:the} we 
theoretically analyze the joint photodetection of classically and quantum
correlated fields. In section~\ref{s:exp} we present the experimental
results obtained for a quantum (twin beam) and a classical (thermal
light divided by a beam splitter) light. In Section~\ref{s:disc} we
discuss the experimental results and draw our conclusions in
Section~\ref{s:outro}.
\section{Quantum versus classical correlations}\label{s:the}
Our aim is to assess the use of joint photodetection as a method to
discriminate classical correlations from entanglement. The scheme we are
going to consider is the following: two modes of radiation, say $\hat
a_1$ and $\hat a_2$, are independently measured by two photodiodes, and
the resulting photocurrents $\hat m_1$ and $\hat m_2$ are then
electronically manipulated and analyzed. In the following we first
investigate the use of the correlation function as an entanglement
marker, and then pass to consider the difference photocurrent, of which
we analyze both the variance and the distribution as a whole. The
different markers are compared in order to discriminate entangled
twin-beam (TWB) of radiation from i) a two-mode factorized coherent
state showing no correlations, and ii) a two-mode thermal beam, showing
classical correlations only. Since entanglement of TWB is a monotone
function of its energy the comparison are performed for fixed mean
number of photons of the involved signals.
\par
Entangled twin-beam $\hat R_X= |X\rangle\rangle\langle\langle X|$ 
are obtained in quantum optics from (spontaneous) parametric
downconversion (SPDC) in second order nonlinear crystals. The expression
in the number basis is given by
\begin{eqnarray}
|X\rangle\rangle = \sqrt{1-|x|^2} \sum_k x^k |k\rangle_1 \otimes
|k \rangle_2  \label{twb}\;,
\end{eqnarray}
where $|k\rangle_j$ denotes a Fock number state in the Hilbert space
of the $j$-th mode. The parameter $x$ satisfies $|x|<1$ and
may be taken as real without loss of generality. The value of $x$
depends on  the crystal length and on the nonlinear susceptibility,
whereas the mean photon number of the TWB is given by $\langle\langle
X|\hat n_1+ \hat n_2|X\rangle\rangle = 2N$, where
$\hat n_j=\hat a_j^\dag \hat a_j$ with $j=1,2$, and $N=x^2/(1-x^2)$
is the mean photon number of each beam.
As a benchmark for uncorrelated classical signals we consider
a two-mode coherent state of the same energy of the TWB, {\em i.e.}
$\hat R_\alpha = |\alpha\rangle_1 {}_1\langle \alpha| \otimes
|\alpha\rangle_2{}_2\langle \alpha| $ with $|\alpha|^2=N$.
On the other hand,  as a reference for classically correlated
signals we consider the state obtained by sending a thermal state
on a balanced beam-splitter whose second port is left unexcited.
In general, if we mix a quantum state $\varrho$ with the vacuum
in a beam-splitter of transmissivity $\tau$, the outgoing state
is described by the density matrix
\begin{eqnarray}
\fl \hat R= \sum_{stpq} \tau^{\frac{s+t}{2}} (1-\tau)^{\frac{p+q}{2}}
\: \varrho_{p+s,t+q}\: \sqrt{
\left( \begin{array}{c}  p+s \\ s \end{array}\right)
\left( \begin{array}{c}  q+t \\ q \end{array}\right)
} \: |s\rangle\langle t| \otimes |p\rangle\langle q|
\label{BsOut}\;
\end{eqnarray}
where $\varrho_{h,k}=\langle h|\hat \varrho |h\rangle$ are the matrix elements
of the input state. In our case $\tau=1/2$ and the input state is a
thermal state with $2N$ mean photon number, {\em i.e}
$\hat \varrho \equiv \hat \nu$ with $\nu_{h,k} = \delta_{h,k}
(1+ 2N)^{-1} [2N/(1+2N)]^k$. We will denote the state obtained
in this way as $\hat R_\nu$. As it can be easily seen by evaluating the
eigenvalues of the partial transpose $\hat R^\theta_\nu$, the state exiting
a beam-splitter fed by a thermal state is never entangled, though it may
show a high degree of  classical correlations. \par
We assume that photodetection is performed with quantum efficiency $\eta$
and no dark counts. The probability operator-valued measure (POVM) of
each detector, describing the statistics of detected photons, is thus given by a
Bernoullian convolution of the ideal number operator spectral measure
$\hat P_{n_j}= |n_j\rangle\langle n_j|$
\begin{eqnarray}
\hat \Pi_{m_j} = \eta_j^{m_j} \sum_{n_j={m_j}}^\infty  (1-\eta_j)^{n_j-m_{j}}
\left( \begin{array}{c}  n_j\\ m_j \end{array}\right)\: \hat P_{n_j}
\label{povm1}\;,
\end{eqnarray}
with $j=1,2$. The joint distribution of detected photons $p(m_1,m_2)$
can be evaluated by tracing over the density matrix of the two modes,
{\em i.e.} $p(m_1,m_2)=\hbox{Tr}\left[\hat R\:\hat \Pi_{m_1}\otimes\hat
\Pi_{m_2} \right]$ while the moments $\langle \widehat{m_1^p}
\widehat{m_2^q} \rangle \equiv
\hbox{Tr}\left[\hat R \:\widehat{m_1^p} \widehat{m_2^q} \right]$
of the distribution are evaluated by means of the operators
\begin{eqnarray}
\widehat{m_j^p} = \sum_{m_j} m_j^p \: \hat \Pi_{m_j} = \sum_{n_j=0}^\infty
(1-\eta_j)^n \:G_{\eta_j}(n_j)\: \hat P_{n_j}
\label{moments}\;,
\end{eqnarray}
where
\begin{eqnarray}
G_\eta(n)= \sum_{m=0}^{n}
\left( \begin{array}{c}  n\\ m \end{array}\right)\:
\left(\frac{\eta}{1-\eta}\right)^{m}\!\! m^p \:.
\end{eqnarray}
Of course, since they are operatorial moments of a POVM, we have,
in general, $\widehat{m_j^p} \neq \hat m_j^p$.
The first two moments correspond to the operators
\begin{eqnarray}
\hat m_j &=&\eta _j\hat n_j \nonumber \\
\widehat{m_j^2} &=&\eta^2_j\hat n_j^2 + \eta_j (1-\eta_j) \hat n_j
\label{mom12}\;.
\end{eqnarray}
As a consequence, the variances of the two photocurrents are
larger than the corresponding photon number variances. We have
\begin{eqnarray}
\label{sig}\;
\sigma^2(m_j)\equiv \langle \widehat{m_j^2}\rangle - \langle
\hat m_j  \rangle^2 = \sigma^2(n_j) + \eta_j (1-\eta_j) \langle
\hat n_j\rangle\:.
\end{eqnarray}
The correlation coefficient is defined as
\begin{eqnarray}
\varepsilon = \frac{\left\langle (\hat m_1 - \langle \hat m_1\rangle)
(\hat m_2 - \langle \hat m_2\rangle)\right\rangle}{\sigma(m_1)\sigma(m_2)}
\label{eps}\;
\end{eqnarray}
where $\hat m_j$ and $\sigma^2(m_j)$ are given in Eqs. (\ref{mom12})
and (\ref{sig}) respectively.  Of course, for factorized
coherent states we have $\varepsilon_\alpha=0$, while for the TWB and the
thermal states we have
\begin{eqnarray}
\varepsilon_X =\frac{(1+N)\sqrt{\eta_1 \eta_2}}{\sqrt{(1+\eta_1 N)(1+\eta_2 N)}}
\qquad
\varepsilon_\nu=\frac{N \sqrt{\eta_1 \eta_2}}{\sqrt{(1+\eta_1 N)(1+\eta_2 N)}}
\label{eps1}\;
\end{eqnarray}
which, for $\eta_1=\eta_2$ reduce to
\begin{eqnarray}
\varepsilon_X =\frac{(1+N)\eta}{1+\eta N} \qquad
\varepsilon_\nu =\frac{N \eta}{1+\eta N}
\label{eps2}\;.
\end{eqnarray}
As it is apparent from Eqs. (\ref{eps1}) and (\ref{eps2}) the
correlation coefficient cannot provide a reliable discrimination
of classical and quantum correlations for a mean number
of photons larger than few units. As a consequence, any imaging
system based on coincidence detection, cannot be improved by 
using entanglement.
\par
\begin{figure}[h]
\includegraphics[width=.4\textwidth]{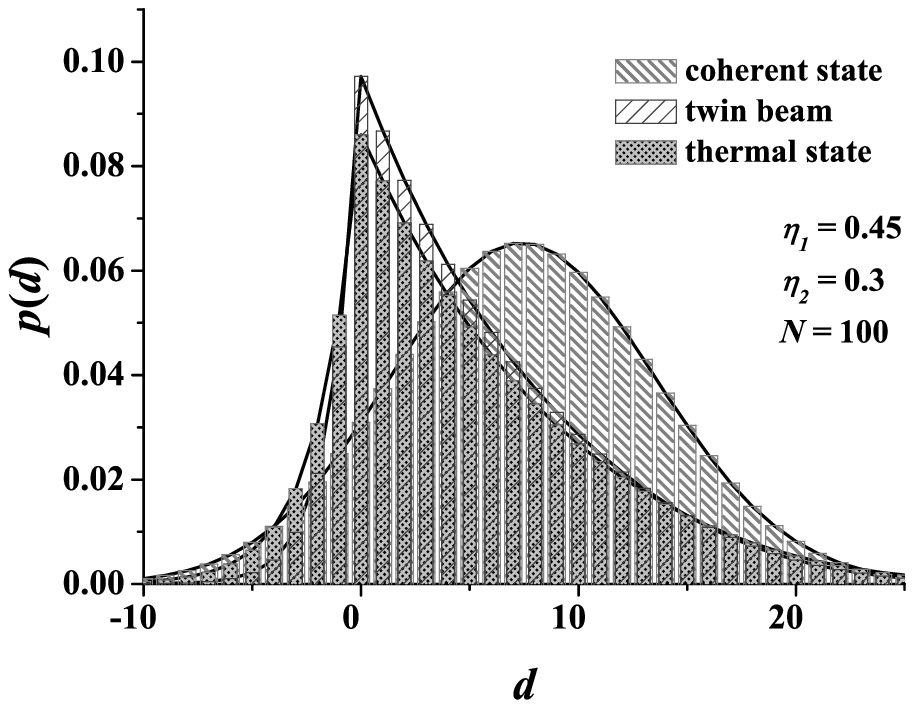}\quad
\includegraphics[width=.4\textwidth]{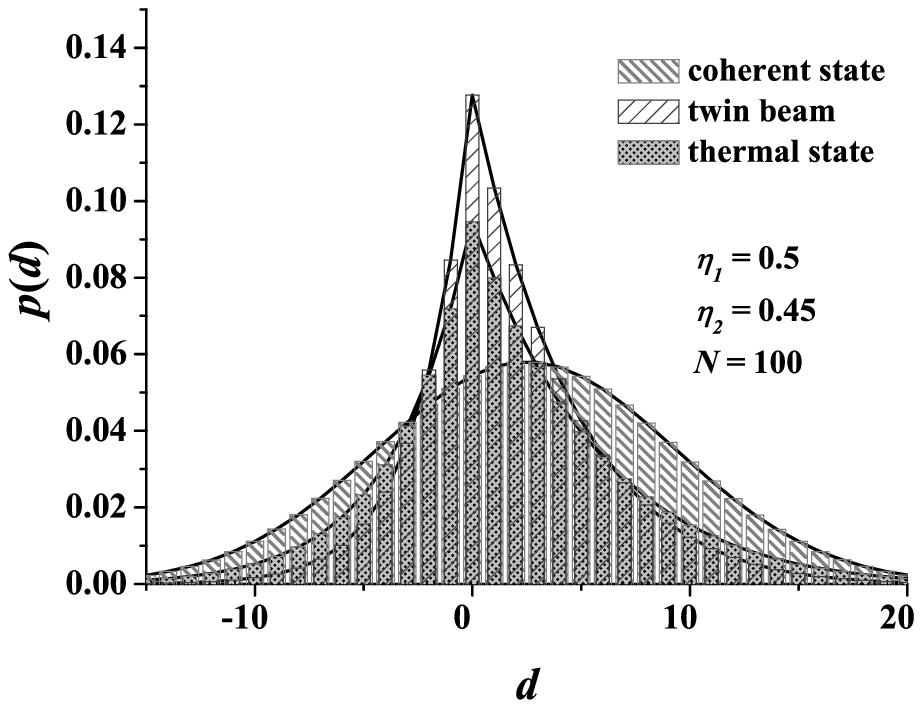} \\
\includegraphics[width=.4\textwidth]{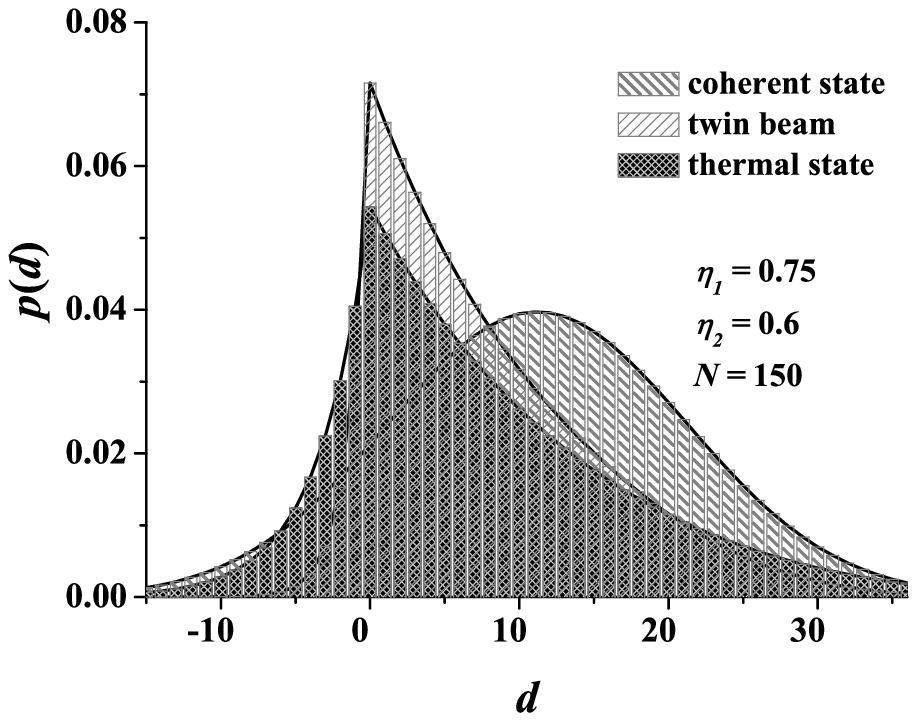}\quad
\includegraphics[width=.4\textwidth]{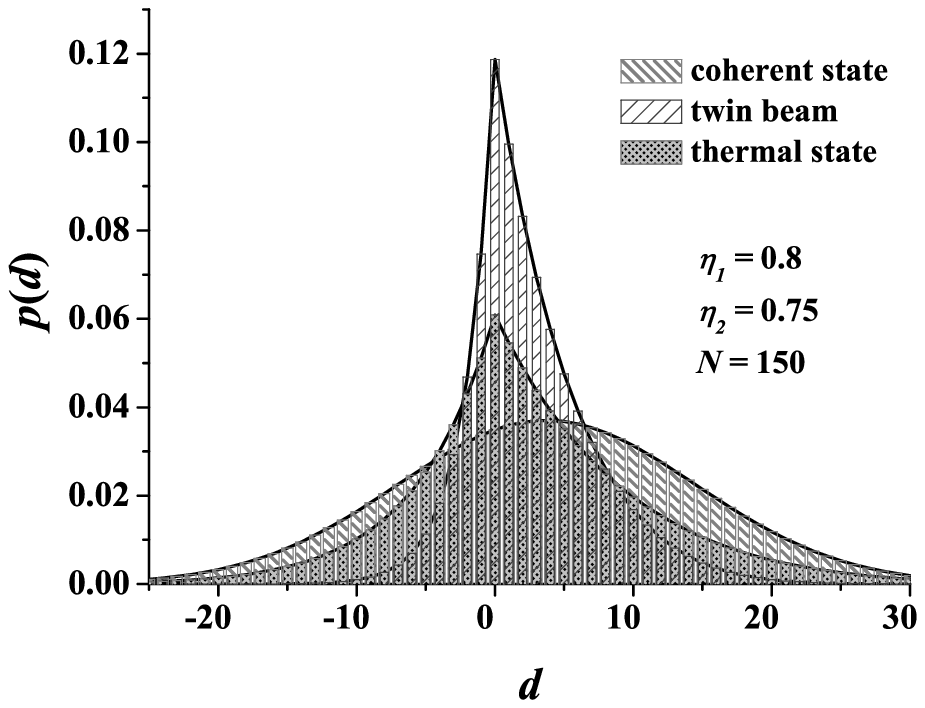} \\
\includegraphics[width=.4\textwidth]{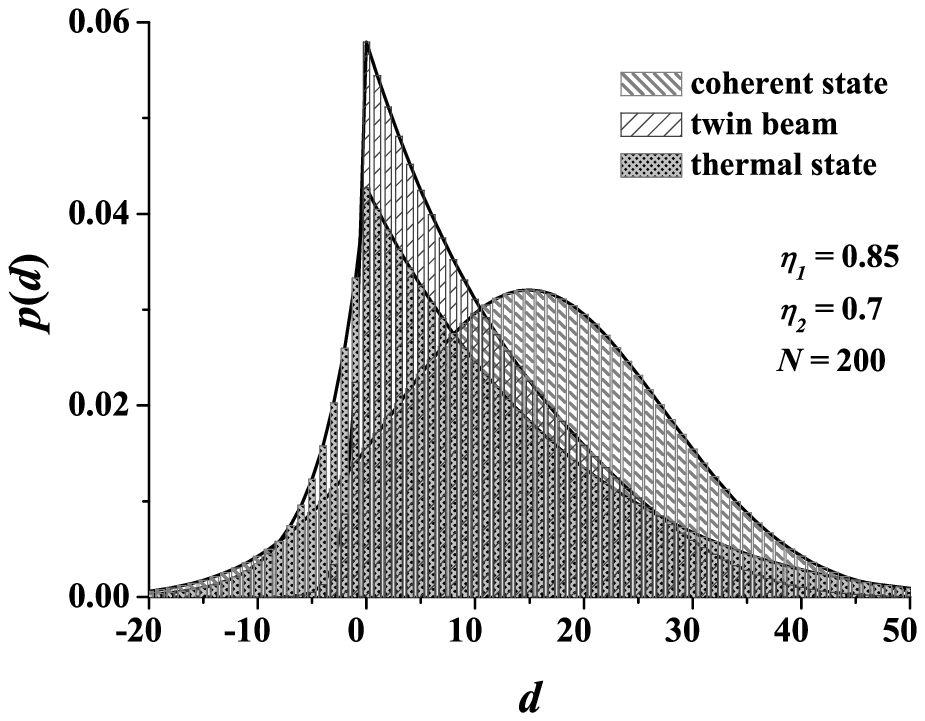}\quad
\includegraphics[width=.4\textwidth]{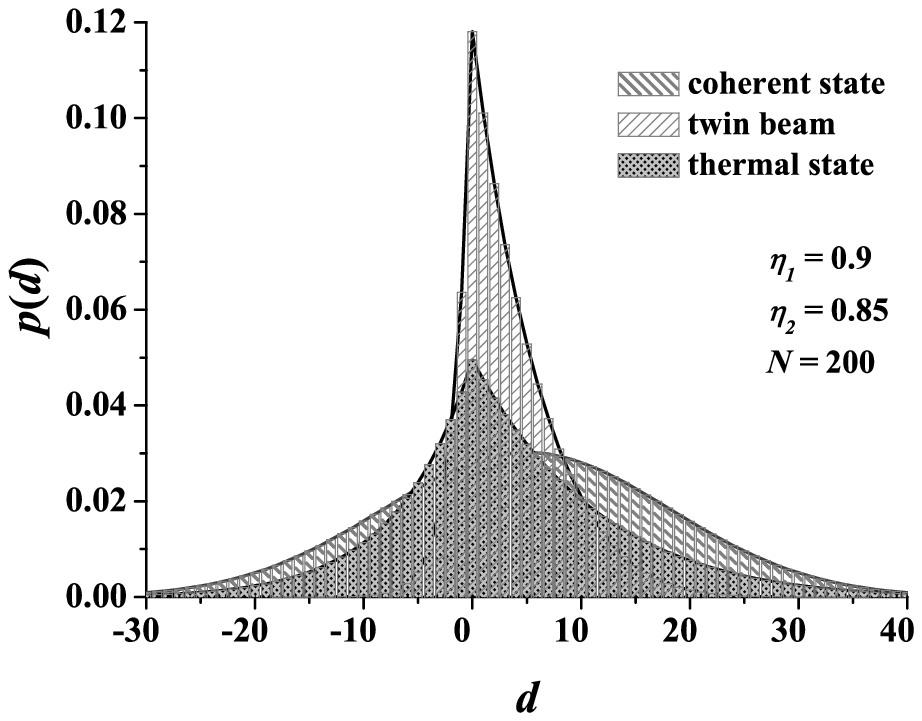} \\
\caption{Probability distributions $p_\alpha(d)$, $p_X(d)$, and
$p_\nu(d)$ for different values of the parameters $\eta_1$, $\eta_2$ and
$N$: the distributions for
a thermal or a coherent state are broader than the corresponding
distribution for the TWB, as far as the quantum efficiencies
are close one each their and their value is not too small.
\label{f:unoT}}
\end{figure}
\begin{figure}[h]
\includegraphics[width=.4\textwidth]{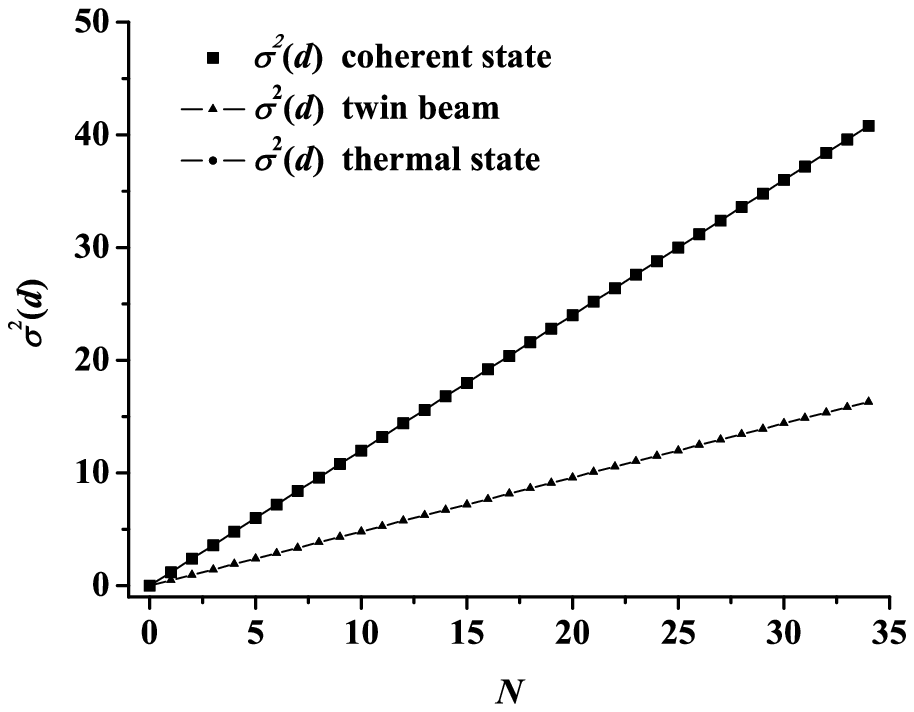} \quad
\includegraphics[width=.4\textwidth]{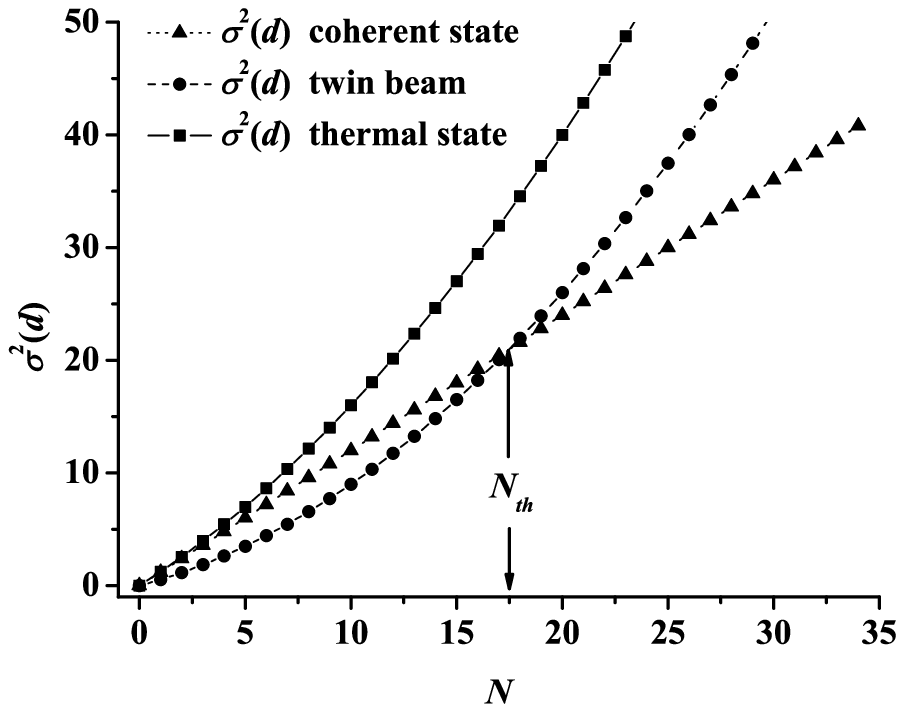}
\caption{Variance $\sigma^2(d)$ of the difference photocurrent as a
function of the mean photon number of the input signal. Left: for
$\eta_1=\eta_2=0.6$; in this case $\sigma^2_X(d)\ll\sigma^2_\alpha(d)
= \sigma^2_\nu(d)$. Right: for $\eta_1=0.5$ and $\eta_2=0.7$; for
different $\eta$'s $\sigma^2_X (d) < \sigma^2_\nu(d)$ and
$\sigma^2_\alpha (d) < \sigma^2_\nu(d)$ $\forall N$, but
$\sigma^2_X (d) < \sigma^2_\alpha (d)$ only for $N < N_{th}
= 2\eta_1 \eta_2/(\eta_1-\eta_2)^2=17.5$.
\label{f:sig1}}
\end{figure}
\begin{figure}[h]
\includegraphics[width=.4\textwidth]{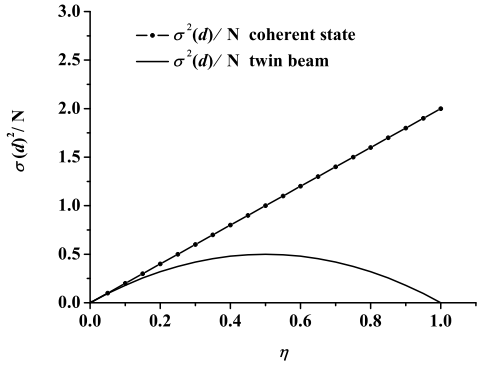}
\caption{Ratio $\sigma^2(d)/N$ between the variance of the
difference photocurrent and the mean photon number of the
signals as a function of the quantum efficiency, assumed to
be equal for the two photodetectors. \label{f:sig2}}
\end{figure}

Let us now consider the quantity obtained by subtracting the
two photocurrents from each other, {\em i.e.} the so-called
difference photocurrent $\hat D=\hat m_1 - \hat m_2$. The statistics of the outcome
can be obtained as $p(d)=\hbox{Tr}\left[\hat R\:\hat \Theta_d\right]$ where
the POVM $\hat \Theta_d$ is given by
\begin{eqnarray}
\hat \Theta_d = \sum_{q=0}^\infty \left\{
\begin{array}{lr}
\hat \Pi_{q+d} \otimes \hat \Pi_q & d>0 \\
\hat \Pi_{q} \otimes \hat \Pi_q & d=0 \\
\hat \Pi_{q} \otimes \hat \Pi_{q+d} & d<0
\end{array}
\right.
\label{povmd}\;,
\end{eqnarray}
with $\hat \Pi_n$ given in Eq. (\ref{povm1}).
The moments of the distribution can be obtained from the operators
\begin{eqnarray}
\hat D = \sum_d d\: \hat \Theta_d = \eta_1 \hat n_1 - \eta_2 \hat n_2
\label{df1}\;,
\end{eqnarray}
\begin{eqnarray}
\widehat{D^2} = \sum_d d^2\: \hat \Theta_d =
(\eta_1 \hat n_1 - \eta_2 \hat n_2)^2
+ \eta_1 (1-\eta_1)\hat  n_1 + \eta_2 (1-\eta_2)\hat  n_2
\label{df2}\;,
\end{eqnarray}
which also provide the variance of the difference photocurrent
$\sigma^2(d)=\langle \widehat{D^2} \rangle - \langle \hat D \rangle^2$.
For the class of states under investigation the difference
photocurrent is distributed as follows
\begin{eqnarray}
\fl p_\alpha (d) = e^{-(\eta_1 + \eta_2) N}\: I_{|d|} (2 N \sqrt{\eta_1
\eta_2}) \: J_{\alpha d} \\
\fl p_X (d) = \frac{1}{1+N} \sum_{n=0}^\infty \sum_{q=n+|d|}^\infty
\left(\frac{\eta_1 \eta_2 N}{1+N}\right)^n
\left( \begin{array}{c}  q\\ n \end{array}\right) \:
\left( \begin{array}{c}  q\\ n+|d| \end{array}\right)
\nonumber \\ \lo \times
\left[(1-\eta_1)(1-\eta_2)\right]^{q-n} \: J_{X d} \\
\fl p_\nu (d) = \frac{1}{1+2 N}\sum_{n=0}^\infty \left(\frac{\eta_1
\eta_2}{(1-\eta_1)(1-\eta_2)}\right)^n \sum_{q, q'} \left(\frac{N}{1+2N}\right)^{q+q'}
\left( \begin{array}{c}  q+q'\\ q \end{array}\right)
\nonumber \\ \lo \times (1-\eta_1)^q (1-\eta_2)^{q'}\:J_{\nu d}
\label{ps}\;,
\end{eqnarray}
where $I_n(x)$ denotes a modified Bessel' function of the first kind,
and the $J$ quantities are given by
\begin{eqnarray}
J_{\alpha d} =
\left\{
\begin{array}{lr}
\left(\frac{\eta_1}{\eta_2}\right)^{d/2} & d\geq 0 \\
\left(\frac{\eta_2}{\eta_1}\right)^{|d|/2} & d\leq 0
\end{array}
\right. \qquad
J_{X d} =
\left\{
\begin{array}{lr}
\left(\frac{\eta_1}{1-\eta_1}\right)^{d} & d\geq 0 \\
\left(\frac{\eta_2}{1-\eta_2}\right)^{|d|} & d\leq 0
\end{array}
\right.
\label{Js}\;, \\
J_{\nu d} =
\left\{
\begin{array}{lr}
\left( \begin{array}{c}  q\\ n+d \end{array}\right)
\left( \begin{array}{c}  q'\\ n \end{array}\right)
\left(\frac{\eta_1}{1-\eta_1}\right)^{d} & d\geq 0 \\
\left( \begin{array}{c}  q'\\ n+|d| \end{array}\right)
\left( \begin{array}{c}  q\\ n \end{array}\right)
\left(\frac{\eta_2}{1-\eta_2}\right)^{|d|} & d\leq 0
\end{array}
\right. \:.
\end{eqnarray}
In Equation (\ref{ps}) the sum are over $q=n+|d|,...$, $q'=n,..$
or $d\geq 0$ and over $q'=n+|d|,...$, $q=n,..$ otherwise.
The distributions are symmetric for $\eta_1=\eta_2$ and
asymmetric otherwise. In Fig. \ref{f:unoT}
we display the distributions $p_\alpha(d)$, $p_X(d)$, and  $p_\nu(d)$
for different values of the parameters $\eta_1$, $\eta_2$ and
$N$. As it is apparent from the plots, the distributions for
a thermal or a coherent state are broader than for the TWB, as far
as the quantum efficiencies are close to each other and their value
is not too small.  In order to quantify this statement more explicitly
we have evaluated, by using Eqs. (\ref{df1}) and (\ref{df2}), the
variance of the difference photocurrent for the three types of states.
We have
\begin{eqnarray}
\label{vara}
\sigma^2_\alpha (d) &=& (\eta_1+\eta_2) N
\stackrel{\eta_1=\eta_2}{\longrightarrow} 2 \eta N\\
\sigma^2_\nu (d) &=& (\eta_1 - \eta_2)^2 N^2 + (\eta_1+\eta_2) N
\label{varn}
\stackrel{\eta_1=\eta_2}{\longrightarrow} 2 \eta N\\
\sigma^2_X (d) &=& (\eta_1 - \eta_2)^2 N^2 + (\eta_1+\eta_2 - 2 \eta_1
\eta_2) N \stackrel{\eta_1=\eta_2}{\longrightarrow} 2 \eta (1-\eta)N
\label{varx}\;.
\end{eqnarray}
For $\eta_1=\eta_2=\eta$ the variances for the two classical states
are equal, and larger than for the TWB state: the difference
being more pronounced the greater is the $\eta$ value. On the other
hand, if the two quantum efficiencies are different, we have
$\sigma^2_\alpha (d) < \sigma^2_\nu (d)$ and
$\sigma^2_X < \sigma^2_\nu (d)$ for
any value of the mean photon number $N$, whereas $\sigma^2_X (d)
< \sigma^2_\alpha (d)$ only for numbers of photon below the
threshold value
\begin{eqnarray}
N_{th}=\frac{2\eta_1 \eta_2}{(\eta_1-\eta_2)^2}
\label{Nth}\;.
\end{eqnarray}
In other words, for equal quantum efficiencies the variance of the
difference photocurrent is a good marker to discriminate between
quantum and classical correlations, whereas for different
quantum efficiencies this statement is true only for signals with
a small number of photons.
In Figs. \ref{f:sig1} we report the variances $\sigma^2(d)$ as 
a function of the mean number of photons for both 
$\eta_1=\eta_2$ and $\eta_1\neq \eta_2$, whereas in Fig. \ref{f:sig2}
we show $\sigma^2(d)/N$ for $\eta_1=\eta_2=\eta$ as a function
of $\eta$.
\par
Let us now consider a situation in which the two beams under
investigation contain more than two, say $2\mu$, modes of the
field, while the correlations to be discriminated are still
pairwise. This is a common situation in pulsed experiments where
several temporal modes are simultaneously matched in SPDC, and
are present in thermal beams as well. We assume that the modes
are equally populated. The statistics of counts for each detector
is described by a multimode POVM  of the form
\begin{eqnarray}
\hat Q_m = \otimes_{s=1}^\mu \: \sum_{m_s=0}^{\infty} \hat \Pi_{m_s} \:
\delta(\sum_s m_s - m)
\label{multi}\;,
\end{eqnarray}
where $\hat \Pi_m$ is the single-mode POVM reported in Eq.
(\ref{povm1}).
The statistics of the difference photocurrent between the two
detectors is described by a $2\mu$-mode POVM of the form
(\ref{povmd}), with $\hat \Pi_n$ replaced by $\hat Q_n$.
\par
Since the modes entering each detector are independent on
each other we have $\langle \hat  m_j\rangle \longrightarrow
\langle \sum_s \hat m_{js} \rangle = \mu \langle \hat m_j\rangle$ and
$\sigma^2 (m_j) \rightarrow \sum_s \sigma^2(m_{js}) = \mu
\sigma^2(m_j)$, $j=1,2$. As a consequence the expressions of the
correlation coefficients are still given by Eqs. (\ref{eps1})
with $N$ that should be meant as the total mean number of photons
of the $\mu$ modes. As concerns the distribution of the difference
photocurrent we have, in terms of the probability density
\begin{eqnarray}
p(d) &=& \sum_n \prod_s \sum_{q_s,r_s} p(q_s,r_s)
\left[
\delta(\sum_s q_s - n - d)
\delta(\sum_s r_s - n) \theta(d) \right. \nonumber \\ &+&\left.
\delta(\sum_s q_s - n)
\delta(\sum_s r_s - n-d) \theta(-d)
\right]
\:,\label{d1}
\end{eqnarray}
where $\theta(x)$ is the Heaviside step function. Notice that in
writing Eq. (\ref{d1}), we have already used the fact that the
correlations are pairwise {\em i.e.} that
$p(q_1,r_1,q_2,r_2,...,q_\mu,r_\mu) = \Pi_s \: p(q_s,r_s)$.
By exploiting the delta functions in (\ref{d1}) we may write
\begin{eqnarray}
 p(d) = \sum_{n=0}^\infty &\:&
\sum_{q_1=0}^{d+n}
\sum_{q_2=0}^{d+n-n_1} ...
\sum_{q_\mu=0}^{d+n-q_1-...-q_{\mu-1}} \nonumber \\ &&
\sum_{r_1=0}^n
\sum_{r_2=0}^{n-r_1} ...
\sum_{r_\mu=0}^{n-r_1-...-r_{\mu-1}}
\!\!\!\!\! p(q_1,r_1)  p(q_2,r_2) ...  p(q_\mu,r_\mu)
\end{eqnarray}
for $d\geq 0$ and an analogue expression (with $q_s \leftrightarrow r_s$)
for $d<0$.
\section{Experimentals}\label{s:exp}
We verified the validity of the theoretical analysis on both quantum
and classically correlated light.
\subsection{Twin Beam}\label{ss:expTWB}
\begin{figure}[b]
\includegraphics[width=.4\textwidth,angle=-90]{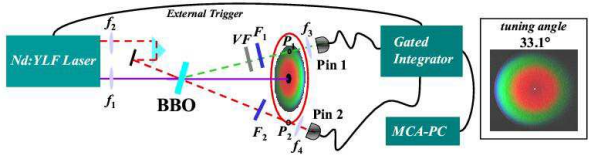}
\caption{Experimental setup for measurements on the TWB: 
BBO, nonlinear crystal; $f_{1-4}$,
lenses; $F_{1-3}$, cut-off filters; $VF$, variable neutral filter;
Pin$_{1-3}$, p-i-n photodiodes; P$_{1,2}, pin-holes$; MCA-PC,
multi-channel analyzer and data acquisition system. Inset: visible
part of the downconversion cones.} \label{f:setupTWB}
\end{figure}

\begin{figure}[h]
\includegraphics[width=.2\textwidth,angle=-90]{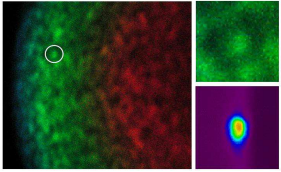}
\caption{Left: Single shot picture of a portion of the signal cone.
Right-top: magnification of a single coherence area around
$\lambda_1$. Right-bottom: intensity map of a typical coherence area
used to estimate its dimensions.} \label{f:coherTWB}
\end{figure}

The quantum state of light we consider is a pulsed twin-beam generated
by a traveling-wave amplifier in non-degenerate configuration. The
layout of the experiment is depicted in Fig.~\ref{f:setupTWB}. As
the pump source we use a frequency-tripled continuous-wave
mode-locked Nd:YLF laser regeneratively amplified at a repetition
rate of 500~Hz (High Q Laser Production, Hohenems, Austria). The
laser delivers $\sim$7.7~ps pulses at the fundamental frequency and
$\sim$4.5~ps pulses at the third harmonics. We obtain intense
spontaneous parametric generation in broadly tunable cones by
injecting the pump field ($\lambda_p = 349\ \mathrm{nm}$) into an
uncoated $\beta$-BaB$_2$O$_4$ crystal (BBO, Fujian Castech Crystals,
Fuzhou, China) cut for type I interaction (cut angle: 34 deg) having
$10\times10$ mm$^2$ cross-section and 4 mm thickness. The pump beam,
which emerges from the laser slightly divergent, is focused by lens
$f_1$ of 50~cm focal length. The crystal tuning angle is $33.1\
\mathrm{deg}$ and the visible portion of the cones projected on a
screen beyond the BBO is shown in the inset of
Fig.~\ref{f:setupTWB}.
We operate in a dichromatic configuration by choosing the frequency
of the laser second harmonics ($\lambda_1 = 523\ \mathrm{nm}$) for
the signal and consequently the frequency of the laser fundamental
($\lambda_2 = 1047\ \mathrm{nm}$) for the idler. For alignment
purposes, a portion of the fundamental beam emerging from the laser
is injected in the crystal together with the pump beam so as to
obtain a well recognizable spot of amplified seeded down conversion.
The selection of the two components of the twin beam is performed by
means of two pin-holes, $P_1$ and $P_2$, having suitable dimensions,
located on the outputs of the seeded process. In order to decide the
dimensions of the pin-holes, such to collect a single coherence area at a
time, we have to determine the dimensions of
the coherence areas of the generated fields.
In Fig.~\ref{f:coherTWB} (left) we show the single-shot picture of a
portion of the signal cone taken with a digital camera (model
Coolpix 990, Nikon, resolution $1024\times 768$), in which we can
clearly distinguish the presence of the coherence areas. In the
right part (top) of the same figure we show a magnified single
coherence area around $\lambda_1$ (green light) and (bottom) the 
intensity map of a typical coherence area taken with a CCD camera (model TM-6CN,
Pulnix, operated at high-resolution). It is easy to demonstrate that
the dimensions of the coherence areas in the idler beam (IR)
corresponding to the measured signal beam scale according to the
ratio of the involved wavelengths so that the dimensions for the
idler are doubled with respect to the signal \cite{Agliati}.
Accordingly, as shown in Fig.~\ref{f:setupTWB}, to select a 
single coherence area on signal and idler, we locate two pin-holes 
(diameter $\simeq 3.5\ \mathrm{mm}$, on the signal and diameter $\simeq 7\
\mathrm{mm}$, on the idler) at a distance of $72.5\ \mathrm{cm}$
from BBO. The light selected by the pin-holes is then focused with
two lenses ($f_3$ and $f_4$, focal length 25 mm) on two p-i-n
photodiodes (Si 85973-02 Hamamatsu, 1 ns time-response, 500 $\mu$m
diameter sensitive area on the green and InGaAs G8376-05, Hamamatsu,
5 ns time-response, 500 $\mu$m diameter sensitive area on the IR)
having nominal quantum efficiencies $\eta_1= 0.92$ and $\eta_{2} =
0.78$ respectively. The current outputs of the photodiodes are
integrated over a synchronous gate of suitable time duration (40 ns)
by a boxcar averager that is operated as gated integrator in external
trigger modality. The boxcar output is digitized by a 13-bit
converter (SR250, Stanford Research Systems, with 50 mV full-scale)
and the counts stored in a PC based multi channel analyzer (MCA).
The measurements are performed by inserting a variable filter ($VF$
in the figure) in front of the photodiode detecting the signal, and
by carefully adjusting it to balance the quantum efficiencies of the
two detection branches of the setup. The interpretation of the
output data must take into account the presence of cut-off filters,
inserted to eliminate residual pump and all stray light, the overall
quantum efficiency of the detection apparatus results to be $\eta_1
\simeq \eta_2 = 0.67$. We verify the linearity of the boxcar
integrators and measure the conversion coefficients ($\alpha_1
=6.7182\times 10^{-8}\ \mathrm{V}$ and $\alpha_2 =8.3043\times
10^{-8}\ \mathrm{V}$) by linking the voltage output of the
digitizer to the number of electrons forming the photocurrent output
pulse of the detectors at each laser shot. The relations among the
statistics of the number of photons incident on the detector,
$p_{ph}(n)$, the statistics of the number of detected photons,
$p_{el}(m)$, and the statistics of the output voltages of the
acquisition apparatus, $p_{out}(v)$, are given by
\begin{eqnarray}
p_{el}(m)&=&\sum_{n=m}^{\infty} \left(
\begin{array}{c}n\\m\end{array}\right)
\eta^m (1-\eta)^{n-m} p_{ph}(n)\\
p_{out}(v) &=& C p_{el}(\alpha m)\ ,\label{eq:bern}
\end{eqnarray}
being $\alpha$ the measured conversion coefficient mentioned above
and $C$ a normalization coefficient. If we limit our analysis to the
first two moments of the distributions, the experimental outputs are
linked to Eq.s~(\ref{mom12}) and (\ref{sig}) by
\begin{eqnarray}
&&V=\alpha M=\alpha\eta N\\
&&\sigma_{out}^2(v) = \alpha^2 \sigma_{el}^2(m) =
\alpha^2\left[\eta^2 \sigma_{ph}^2(n)+\eta(1-\eta)N\right]\
,\label{eq:moments}
\end{eqnarray}
where for the sake of clarity we have defined
$\sigma_{el}^2(m)\equiv \sigma^2(m)$ and $\sigma_{ph}^2(n)\equiv
\sigma^2(n)$ [see Eq.~(\ref{sig})]. Note that in general the
statistical distribution for the measured outputs is different from
that of the incident photons. However, in both our cases (quantum and
classical), the statistical distributions of the detected photons and 
of the voltage outputs are thermal ones.
\par
\begin{figure}[h]
\includegraphics[width=.4\textwidth]{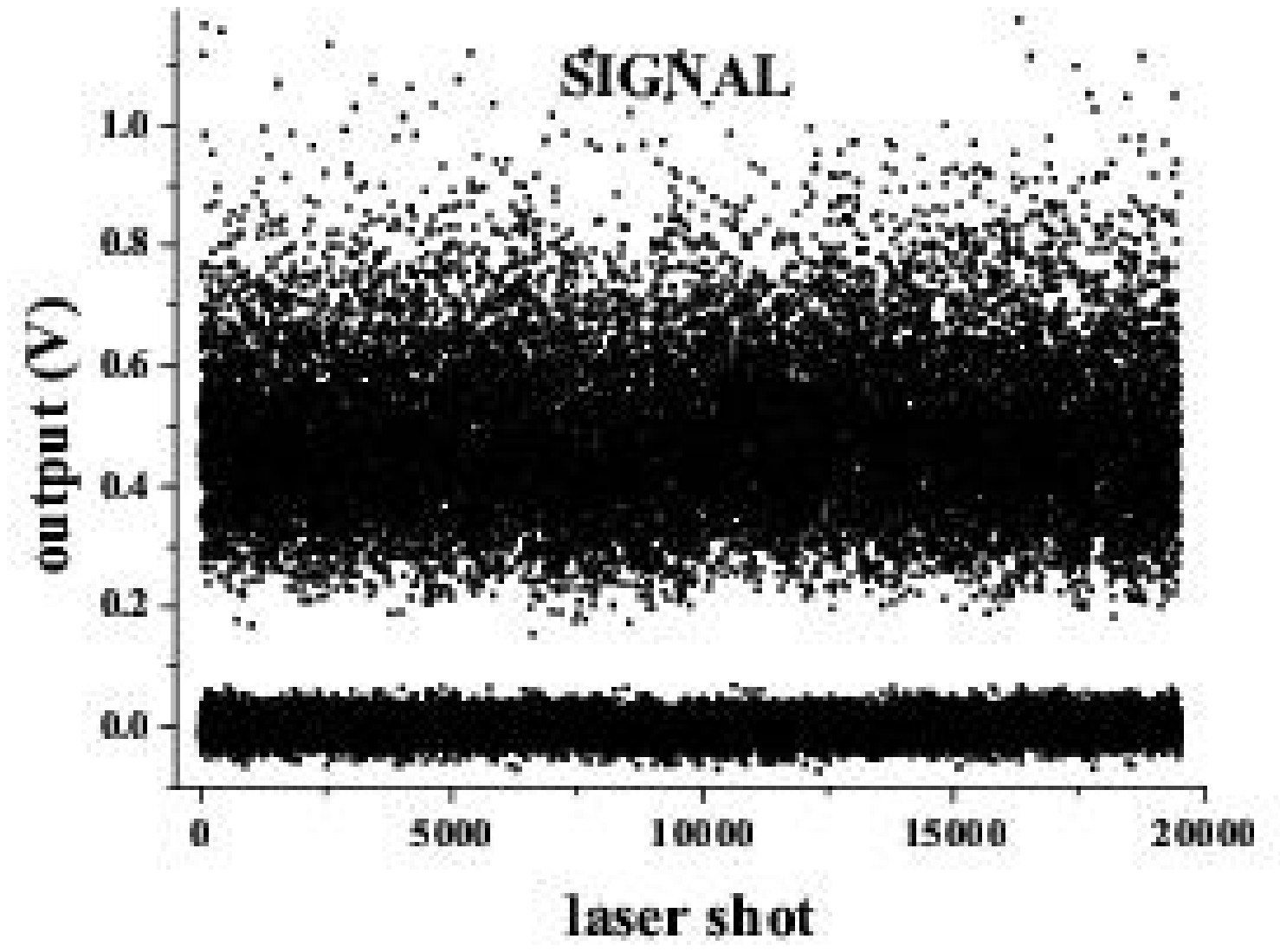}\hspace{5mm}
\includegraphics[width=.4\textwidth]{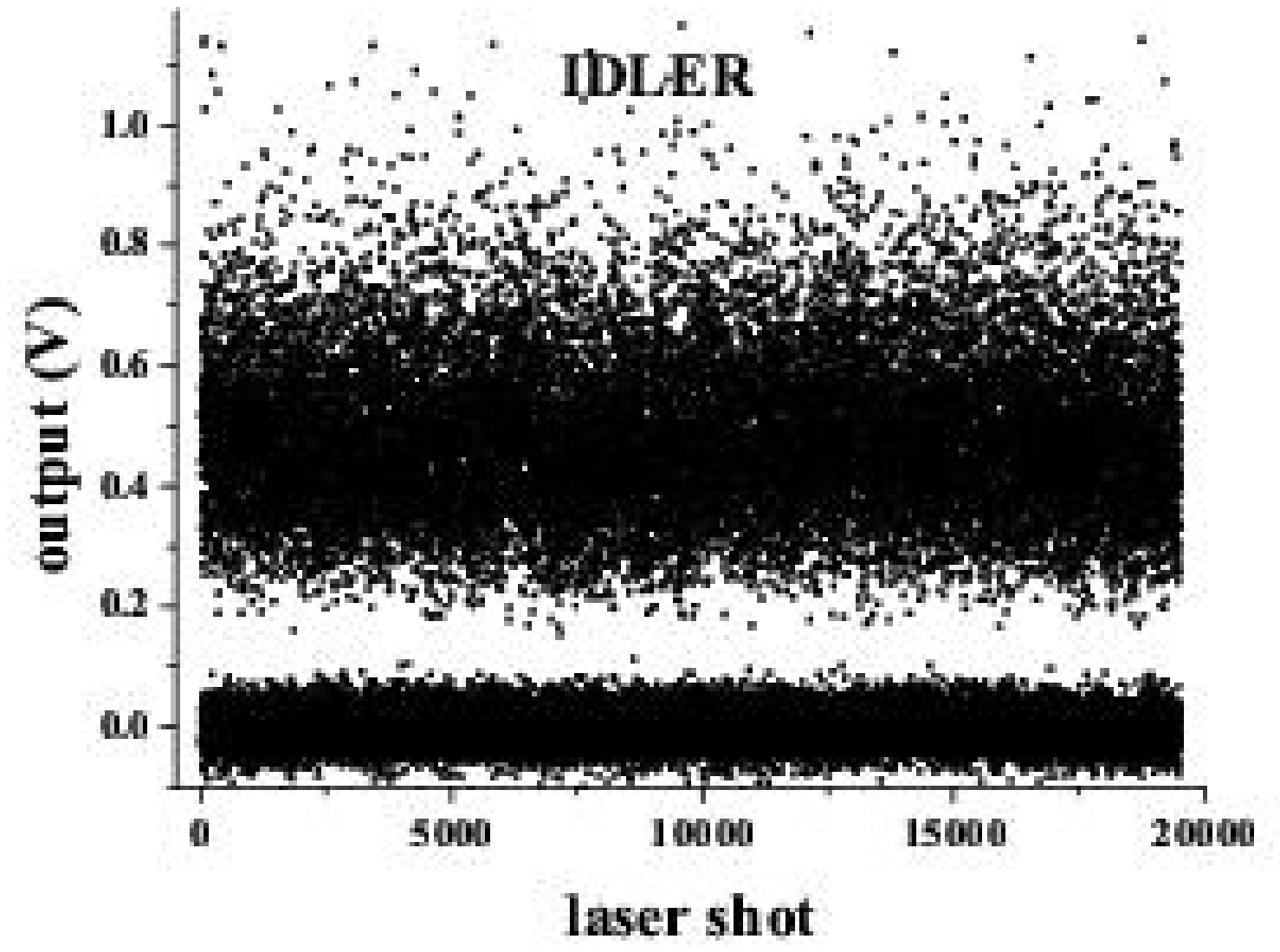}
\caption{Left: Voltage outputs for the signal beam at $\lambda_1 = 523$
nm for a sequence of laser shots and noise. Right: Voltage outputs
for the idler beam at $\lambda_2 = 1047$ nm for the same sequence of
independently.} \label{f:tracesX}
\end{figure}
\begin{figure}[h]
\includegraphics[width=.4\textwidth]{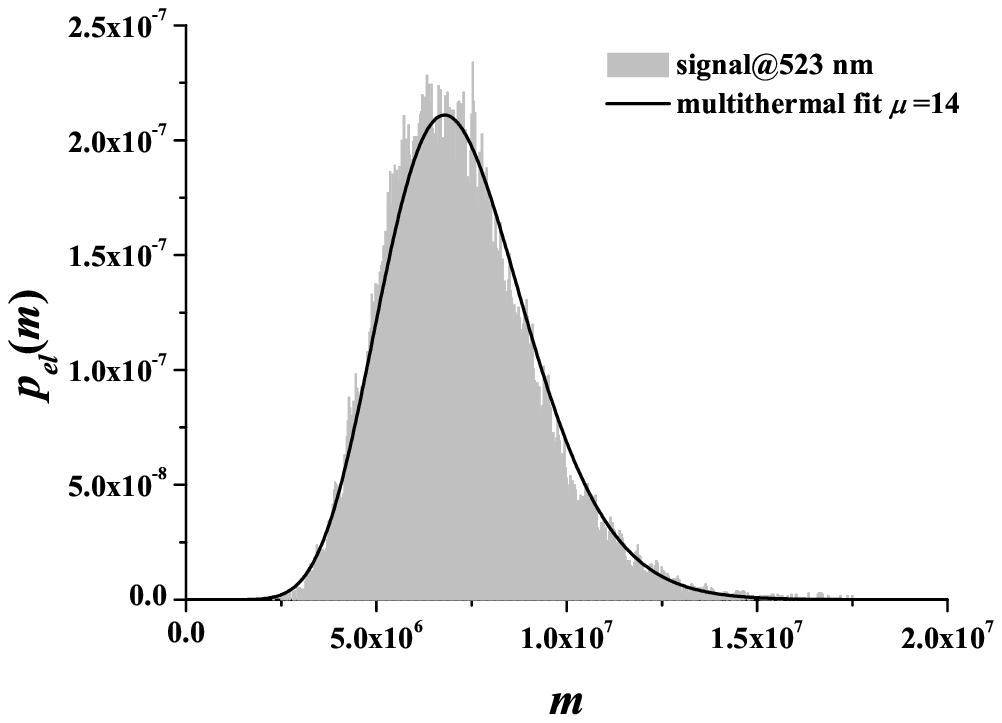}\hspace{5mm}
\includegraphics[width=.4\textwidth]{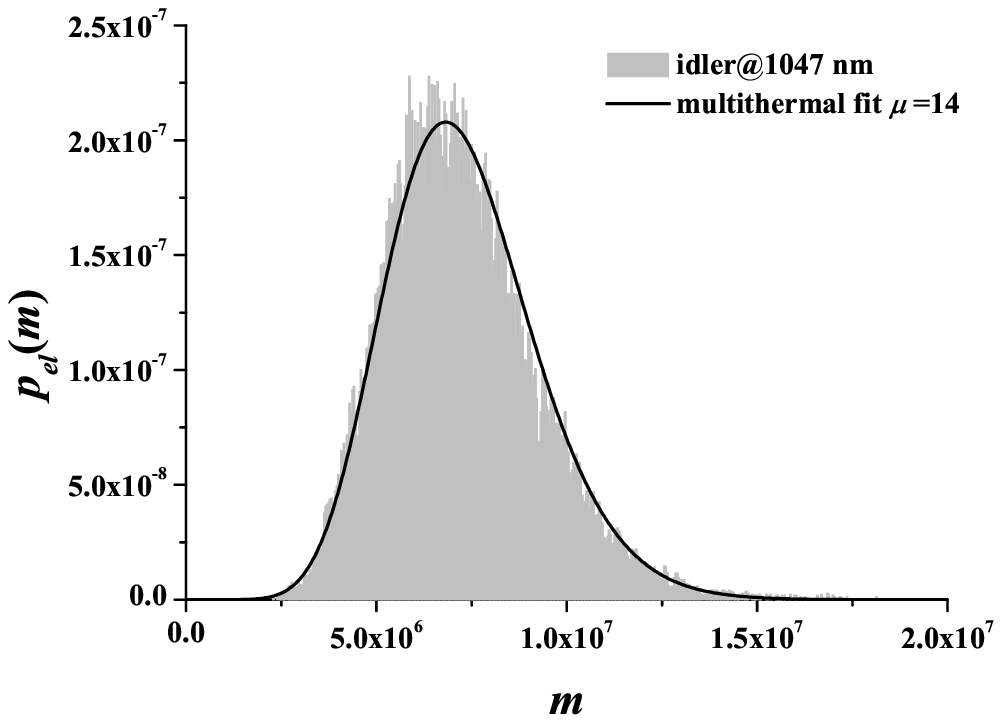}
\caption{Left: histogram of the intensity distribution of 
signal beam output at $\lambda_1 = 523$ together with the corresponding 
multithermal fit. Right: histogram of the intensity distribution of 
idler beam output at $\lambda_2 = 1047$ nm together with the corresponding 
multithermal fit .} \label{f:statX}
\end{figure}
\begin{figure}[h]
\includegraphics[width=.4\textwidth]{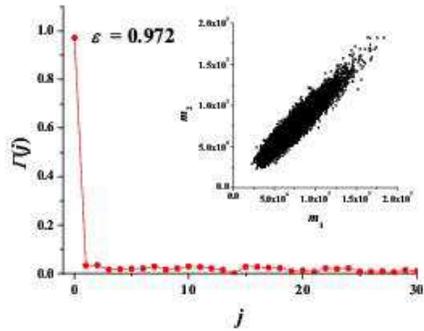}
\caption{Correlation coefficient between signal and idler beam 
as a function of the delay in the laser shot. Inset: values of 
the detected photons in the idler as a function of those in the 
signal in each laser shot.} \label{f:correlX}
\end{figure}

In Fig.~\ref{f:tracesX} we show the recorded signal (left) and idler
(right) outputs of the photodiodes as a function of the laser shot,
together with the noise of the detectors. In Fig.~\ref{f:statX} we
the corresponding normalized probability distributions are reported for 
the same data. By looking at the probability distributions in Fig.~\ref{f:statX}
we note that the statistics of the outputs are well fitted by
multithermal distributions \cite{Paleari}, that is the distributions
obtained by the convolution of $\mu$ equally populated thermal modes
\begin{eqnarray}
p_{out,\mu}(v) = \frac{\exp\left(-v\mu/
V_T\right)}{\left(\mu-1\right)!}\times\frac{v^{\mu-1}}{{\left(
V_T/\mu\right)}^\mu}\ , \label{eq:multit}
\end{eqnarray}
where $V_T = \alpha M_T$ is the mean output corresponding to the
overall detected photons mean value $M_T$.
Equation~(\ref{eq:multit}) holds in the high-intensity regime, which
is the present experimental condition. In fact, by using the
measured conversion coefficients on the detection arms of signal and
idler we get $M_1 = 7.225\times 10^6$ and $M_2 = 7.212\times 10^6$
as the mean number of detected photons. As it is well known from the
theory of photodetection \cite{Mandel}, the number of detected modes
can be interpreted as the ratio of the time characteristic of the
measurement (in our case the time duration of the pulse) and the
coherence time characteristic of the field to be measured (in our
case the inverse of the temporal bandwidth of the spontaneous
parametric down conversion) \cite{Paleari}. The continuous lines
superimposed to the histograms of the experimental data in
Fig.~\ref{f:statX} show the convolution integrals, optimized for the
number of temporal modes, of the theoretical distribution in
Eq.~(\ref{eq:multit}) with the system impulse response evaluated
from a measure in the absence of incident light. As expected, the 
signal and idler distributions are well fitted by multithermal
distributions having the same number of modes ($\mu = 14$). Note
that the probability distributions for signal and idler are very
similar to each other. In order to stress the correspondence between
signal and idler, we plot the output of the idler as a function of
that of the signal (see Inset in Fig.~\ref{f:correlX}).
To compare the experimental results with the theoretical
predictions, we first of evaluate the correlation function of the
photocurrents as
\begin{eqnarray}
\Gamma(j)=\frac{\sum_{k=1}^K\left(v_1(k)-\langle
v_1\rangle\right)\left(v_2(k+j)-\langle
v_2\rangle\right)/K}{\sigma(v_1)\sigma(v_2)}\ , \label{eq:correl}
\end{eqnarray}
where the average operations are taken over $K$ (typically
$K=30000$) subsequent laser shots. For $j=0$, Eq.~(\ref{eq:correl})
gives the correlation coefficient $\varepsilon$
\begin{eqnarray}
\varepsilon=\frac{\langle\left(v_1-\langle
v_1\rangle\right)\left(v_2-\langle
v_2\rangle\right)\rangle}{\sigma(v_1)\sigma(v_2)}\ , \label{eq:eps}
\end{eqnarray}
which should be compared with the theoretical predictions of
Eqs.(\ref{eps1}) and (\ref{eps2}).
In Fig.~\ref{f:correlX} we show the correlation coefficient for the data
of Figs.~\ref{f:tracesX} and \ref{f:statX}: the contributions of the noise of
the apparatus, (\textit{i.e.} the variance of the impulse response in
Fig.~\ref{f:statX}), are subtracted from the measured variances of
the experimental data. We get $\varepsilon = 0.97$, to be compared
with a theoretical value of about $1$. Note that subsequent shots
results to be uncorrelated.
\par
\begin{figure}[h]
\includegraphics[width=.4\textwidth]{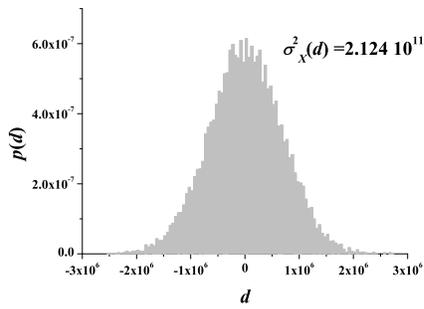}
\caption{Experimental distribution of the difference photocurrent
between signal and idler beams of a TWB.} \label{f:diffTWB}
\end{figure}

As it has been shown in Section \ref{s:the}, the distribution of the
difference photocurrent is a relevant  marker of entanglement. In
Fig.~\ref{f:diffTWB} we plot the distribution of the difference of
the photoelectrons detected on signal and idler, {\em i.e.}
$p(d)=p(m_s-m_i)=p(v_i/\alpha_i-v_s/\alpha_s)$.
The distribution appears almost symmetrical and centered at zero,
which indicates both accurate balance of the detectors' quantum
efficiencies and high correlation in signal/idler photon numbers.
The variance, as evaluated from the data, once the variance of the
noise is subtracted, turns out to be $\sigma^2_X(d)= 2.124\times
10^{11}$.
\subsection{Thermal Light}\label{ss:expTHERM}
\begin{figure}[h]
\includegraphics[width=.4\textwidth,angle = -90]{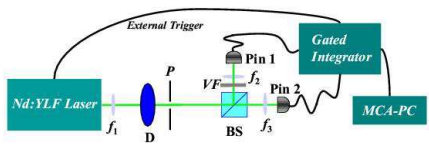}
\caption{Experimental setup for measurements on classically correlated
beams: BS, 50 $\%$ cube beam-splitter; $D$,
moving diffuser; $f_{1-4}$, lenses; $VF$, variable neutral filter;
Pin$_{1-3}$, p-i-n photodiodes; P, iris; MCA-PC, multi-channel
analyzer and data acquisition system. } \label{f:setupCLASS}
\end{figure}

To investigate joint photodetection for 
classically correlated light, we modify the experimental setup 
according to Fig.~\ref{f:setupCLASS}. Pseudo-thermal light has been 
generated by inserting a moving ground-glass diffusing plate in the 
path of the second-harmonics output of the laser ($\lambda = 523$ nm). 
A portion of diffused light is
selected with an iris (in Fig.~\ref{f:setupCLASS}) and then sent
to a 50$\%$ cube beam splitter. The temporal statistics of the
generated light can be described by the same statistics as in
Eq.~\ref{eq:multit} \cite{Goodman}, in which the number of modes can
be varied by changing the dimension of the iris in order to collect
more than one spatial coherence area. The beams emerging from the
beam splitter are then detected by the same apparatus used for the
twin beam, where the pin photodiodes are now identical (model
S3883-02, Hamamatsu, $\eta\simeq 0.71$, nominal) since the two beams
are at the same frequency. The mean number of detected photons on
the two beams are $M_1\simeq M_2\simeq 2.22\times 10^8$.
\par
\begin{figure}[h]
\includegraphics[width=.4\textwidth]{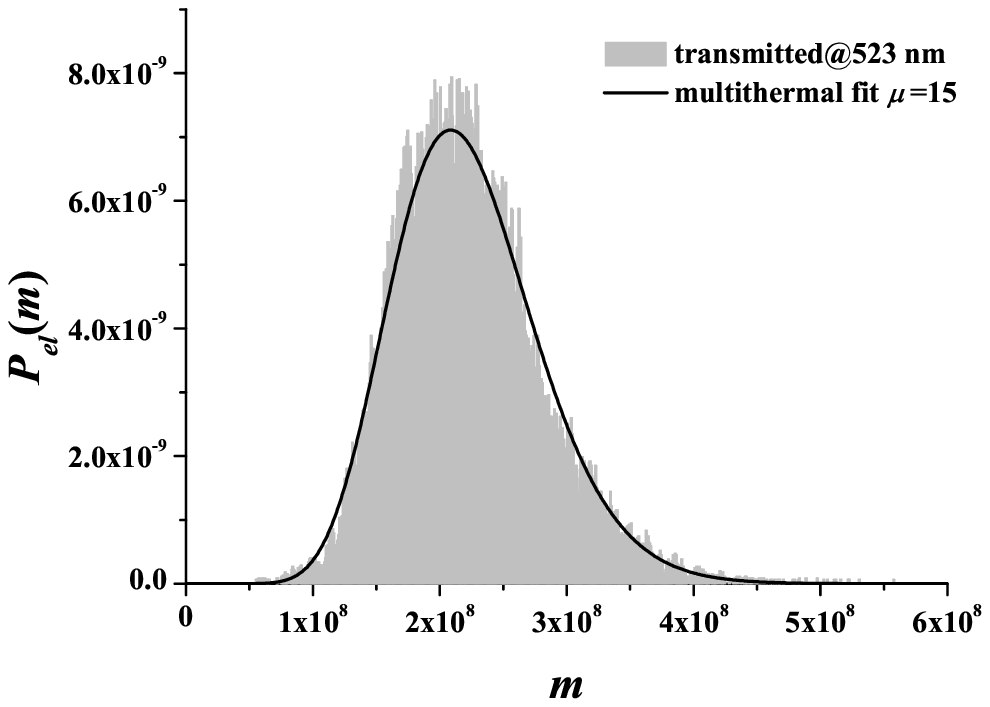}\hspace{5mm}
\includegraphics[width=.4\textwidth]{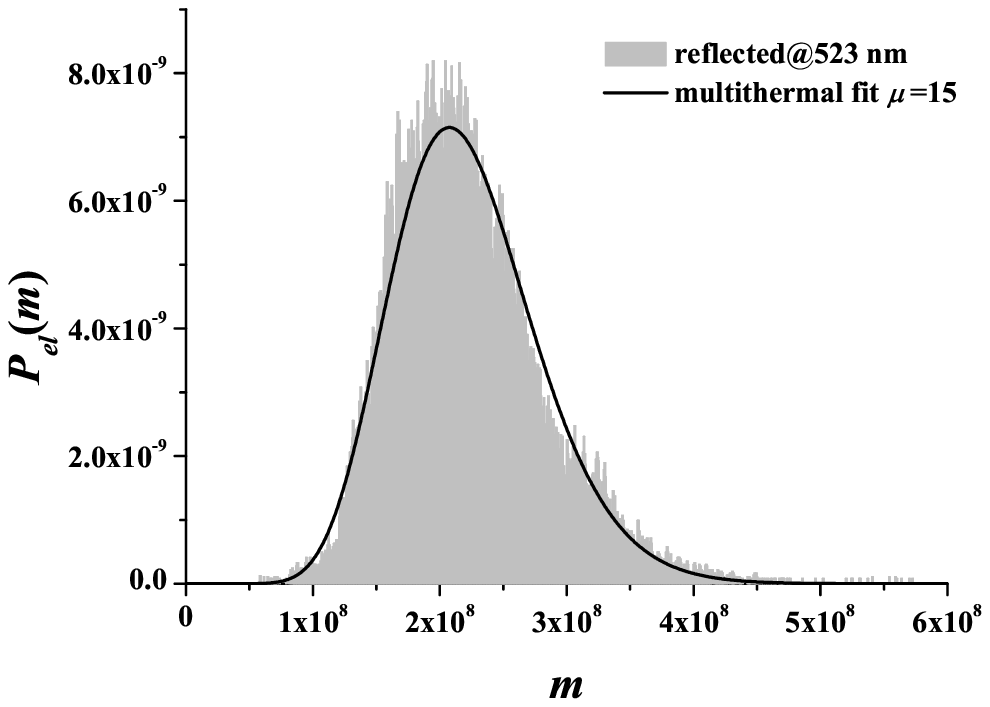}
\caption{Left:histogram of the intensity distribution of 
the transmitted beam together with the corresponding multithermal
fit.Right histogram of the intensity distribution of 
the reflected beam together with the corresponding multithermal
fit.} \label{f:statCLASS}
\end{figure}
\begin{figure}[h]
\includegraphics[width=.4\textwidth]{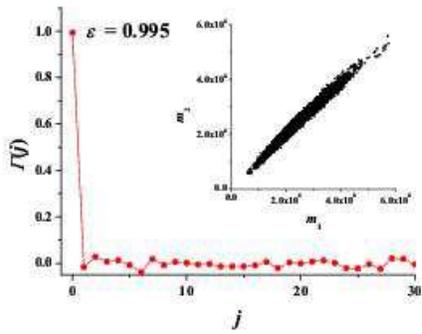}
\caption{Correlation coefficient between transmitted and reflected
beams as a function of the delay in the laser shot. Inset: values of
the detected photons in the reflected beam as a function of those in
the transmitted beam at each laser shot.} \label{f:correlCLASS}
\end{figure}

In Fig.~\ref{f:statCLASS} we show the normalized probability
distributions for the detected photons. The continuous lines
superimposed to experimental data in Fig.~\ref{f:statCLASS} are the
best fits of the data obtained for 15 modes. As in the case of the
twin beam, the two histograms are very similar and suggest a high
degree of correlation that is easily verified evaluating the value
of the correlation function.
In the Inset of Fig.~\ref{f:correlCLASS}, we plot the two voltage
outputs of the beam splitter one \textit{versus} the other, and in
right part the correlation function for the classical beams in which
again the contributions of the noise of the apparatus have been
subtracted from the measured variances of the experimental data. We
get $\varepsilon = 0.995$ to be compared with a theoretical value of
about $1$.
\par
\begin{figure}[h]
\includegraphics[width=.4\textwidth]{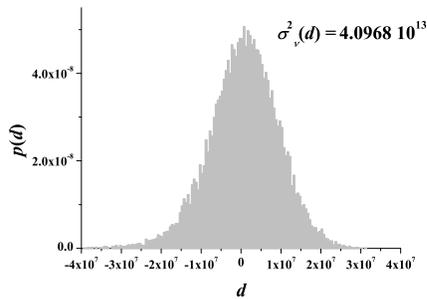}
\caption{Experimental distribution of the difference photocurrent
between the transmitted
and the reflected beams of a pseudo-thermal beam impinging onto a
beam splitter.} \label{f:diffCLASS}
\end{figure}

In Fig.~\ref{f:diffCLASS} we plot the distribution of the difference
of the photoelectrons detected on the two arms of the beam splitter.
Again the distribution appears symmetrical and peaking at zero. The
variance, as evaluated from the data upon subtraction of the noise,
is $\sigma^2_{\nu}(d)= 4.097\times 10^{13}$.
%
\section{Discussion}\label{s:disc}
The experimental results discussed in Section~\ref{s:exp} are
obtained by keeping the values of the quantum efficiencies as close
each other as possible. Therefore, they must be compared with the 
expected values for equal quantum efficiencies and with the 
shot-noise level for the intensities we are working at. The 
theoretical values are $\sigma^2_X(d) = 4.769\times 10^6$ and 
$\sigma^2_\alpha (d) = 1.444\times 10^7$ for the TWB and $\sigma^2_\nu(d)
=\sigma^2_\alpha = 4.446\times 10^8$ for the classically correlated
thermal light. In order to obtain a realistic comparison between theory and
experiment, we have to take into account the presence of noise that
unavoidably affects the experimental data. We identify two main
sources of noise: first of all, the difference between the overall
quantum efficiencies on the two detection branches. In fact,
although the experimental procedure was optimized so as to obtain
the best balanced $\eta$ values, a small residual difference cannot
be excluded, and, as we will see, a small balance error, even local
across the beam to be measured, produces a relevant difference in the
values of $\sigma^2(d)$. On the other hand, we have to take into 
account the unavoidable fluctuations of the laser source which affect 
all the fields under investigation. In fact, the pulsed pump field 
is not a plane wave having constant amplitude. Rather, its statistics 
is more realistically modeled by a Gaussian distribution, \textit{i.e.} 
a Poissonian distribution affected by an excess noise \cite{Loudon}
\begin{equation}
p_p(n) = \frac{1}{\sqrt{2\pi \sigma_p^2}} \exp
\left[-\frac{(n-\langle n_p \rangle)^2}{2\sigma_p^2}\right]\:,
\label{eq:gauss}
\end{equation}
where $\sigma_p^2 = \langle n_p \rangle+\delta_{noise}^2$ and
$\delta_{noise}^2 = x^2 \langle n_p \rangle^2$  is the increase
of the variance due to fluctuations; the quantity $x$ measures the
amount of such a deviation. We will evaluate the influence on the
generated beams of the excess noise in the pump by evaluating the
error propagation.
\subsection{Imbalance of the quantum efficiencies}
To evaluate the modifications of the experimental results due to
imbalance in the quantum efficiencies of the two branches, we equate
the experimental results for $\sigma^2(d)$ with the theoretical
predictions for unbalanced quantum efficiencies of the
photodetectors (see Eqs.~(\ref{varn}) and (\ref{varx}) for
$\eta_1\neq \eta_2$). In the case of TWB, we obtain $0.12\le\vert
\eta_1-\eta_2\vert\le 0.22$ and in the case of classical field
$0.05\le\vert \eta_1-\eta_2\vert\le 0.12$. These values are too
large to be reconciled with the high symmetry of the measured $p(d)$
(see Fig.~\ref{f:diffCLASS}). We can thus conclude that simply
including a difference in the quantum efficiencies on the two
detection branches is not sufficient to account for the experimental
data.
\subsection{Fluctuations in the laser source}
We evaluate the influence of the excess noise of the third-harmonics
pump pulse on the generated beams.
\par
\begin{figure}[h]
\includegraphics[width=.42\textwidth]{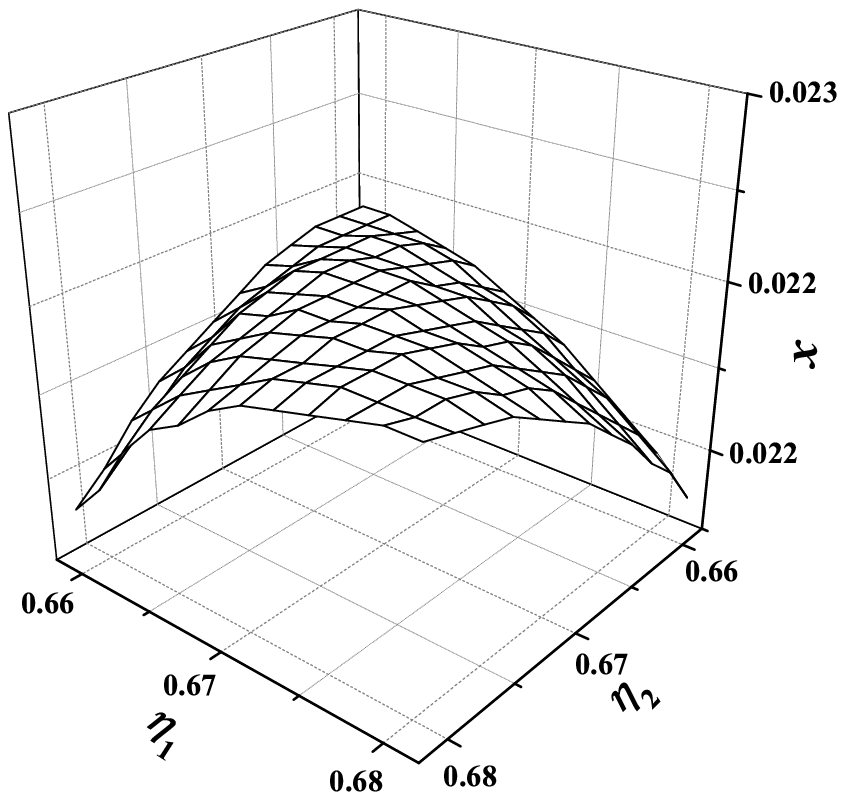}\hspace{1mm}
\includegraphics[width=.42\textwidth]{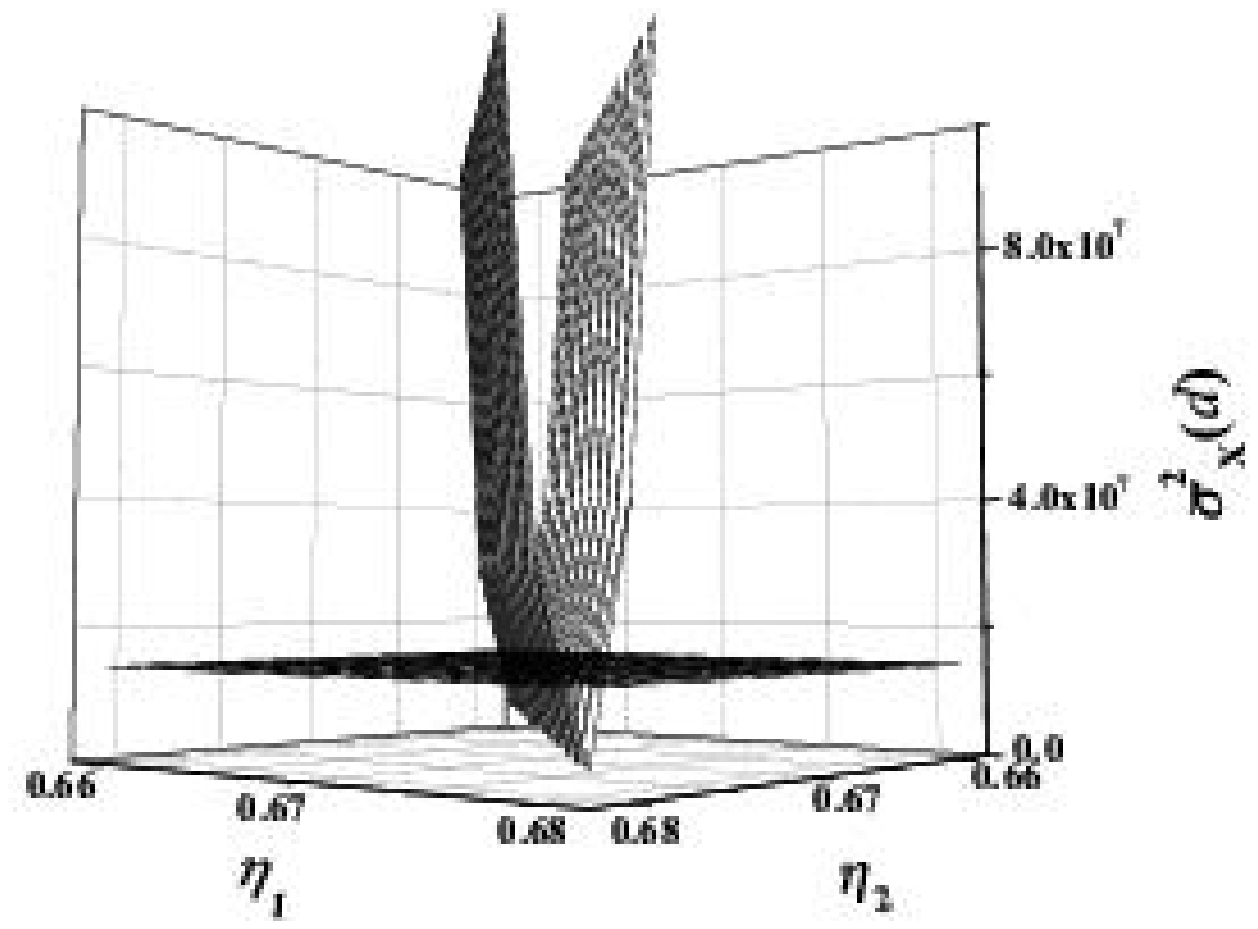}
\caption{Laser fluctuations in experiments with TWB. 
Left: the amount of laser fluctuations $x$ as a function 
of the quantum efficiencies $\eta_1$ and $\eta_2$. Right: values of 
the corrected variance $\overline{\sigma}^2_{X}(d)$ as a function 
of the quantum efficiencies $\eta_1$ and $\eta_2$; the
plane represents the shot-noise value.} \label{f:xP}
\end{figure}

Starting with the SPDC, we recall that the mean photon number in
each component of the generated twin beam is given by
\begin{equation}
N_{X}=\sinh^2(ga_pL)\ , \label{eq:Ntwb}
\end{equation}
where $g$ is a coupling constant, $L$ is the interaction length
inside the crystal and $a_p = \sqrt{N_p/(A_p \tau_p)}$, being $N_p$
the mean photon number, $A_p$ the cross section and $\tau_p$ the
temporal duration of the pump pulse. By applying the
error-propagation theory to Eq.~(\ref{eq:Ntwb}), we get for the
excess noise in the single mode of signal (idler):
\begin{eqnarray}
\delta^2_{X}(n)&=&\sigma^2_p\left(\frac{\partial N_{X}}{\partial
N_p}\right)^2=\left(N_p+x^2
N_p^2\right)\frac{g^2L^2}{A_p\tau_p}\frac{N_{X}^2}{N_p}\nonumber\\
&=& \left(\frac{1}{N_p}+x^2\right)N_{X}^2 \mathrm{arcsinh}^2\sqrt{
N_{X}}\nonumber\\
&\simeq& x^2 N_{X}^2\mathrm{arcsinh}^2\sqrt{N_{X}}\ ,
\label{eq:DeltaNtwb}
\end{eqnarray}
where we used Eq.~(\ref{eq:Ntwb}) and the final approximation holds
for $N_p\gg 1$. In the case of a multithermal beam composed by $\mu$
modes, Eq.~(\ref{eq:DeltaNtwb}) becomes:
\begin{eqnarray}
\delta^2_{X}(n)&=&\frac{N_{X}^2}{\mu}x^2
\mathrm{arcsinh}^2\sqrt{\frac{N_{X}}{\mu}}\ .
\label{eq:DeltaNtwb2}
\end{eqnarray}
The variance of the difference photocurrent can thus be corrected as
\begin{eqnarray}
\overline{\sigma}^2_{X}(d)&=&\sigma^2_{X,sp}(d)-\frac{M_{1}^2}{\eta_1^2\mu}x^2
\mathrm{arcsinh}^2\sqrt{\frac{M_{1}}{\eta_1\mu}}-\frac{M_{2}^2}{\eta_2^2\mu}x^2
\mathrm{arcsinh}^2\sqrt{\frac{M_{2}}{\eta_2\mu}}\ ,
\label{eq:sigmaCORR}
\end{eqnarray}
which is a function of the parameter $x$. We now evaluate the amount
of laser fluctuations (\textit{i.e.} the value of $x$) needed to
reproduce the experimental data. To this aim, we equate
Eq.~(\ref{eq:sigmaCORR}) to Eq.~(\ref{varx}), modified to consider
the presence of $\mu$ modes in the measured field
\begin{eqnarray}
\overline{\sigma}^2_{X}(d) = (\eta_1 - \eta_2)^2
\frac{M_{1,2}^2}{\eta_{1,2}^2\mu} + (\eta_1+\eta_2 - 2 \eta_1
\eta_2) \frac{M_{1,2}}{\eta_{1,2}} \label{varsE1E2}\;,
\end{eqnarray}
and study the dependence of $x$ on the value of the overall quantum
efficiencies on the two detected fields. Notice that, from the
experimental point of view, we have two possible choices for the
value of $N$ appearing in the theoretical formula, namely
$N=M_j/\eta_j$ with $j=1,2$ indicating either signal or idler, in
our experimental conditions $M_1\simeq M_2$ and the two conditions
give very similar results.
Figure~\ref{f:xP} displays the values of $x$ as a function of
$\eta_1$ and $\eta_2$ (left), and the corresponding values of the
corrected $\overline{\sigma}^2_{X}(d)$ as calculated from
Eq.~(\ref{eq:sigmaCORR}) (right). The horizontal plane in on the
right represents the shot-noise level of the measure as calculated
from Eq.~(\ref{vara}). Starting from data in Fig.~\ref{f:xP} we can 
draw two conclusions: On one hand, the experimental data corresponds 
to an amount of laser excess noise equal to $x\simeq 2.24\%$, which 
is compatible with the fluctuations of a pulsed laser. On the other
hand, we have that at the intensities used in our experiments 
we cannot reliably discriminate the measured $\sigma^2_{X,sp}(d)$ 
from the shot-noise level. In fact, the right part of Fig. \ref{f:xP} 
shows that a slight indetermination in the quantum efficiencies may
considerably increase the variance above the shot noise level.
Note that the inclusion of an added noise does not imply a significant 
modification of the variance of the beams, as the total variance of 
signal/idler can be written as
\begin{eqnarray}
\bar\sigma^2_{X} =\frac{N_{X}^2}{\mu}\left(1+x^2
\mathrm{arcsinh}^2\sqrt{\frac{N_{X}}{\mu}}\right)\ .
\label{eq:DeltavarTWB}
\end{eqnarray}
As the correction to unity is less than $3\%$, the measured
distributions are still well fitted by the expected multithermal
distributions.
\par
\begin{figure}[h]
\includegraphics[width=.42\textwidth]{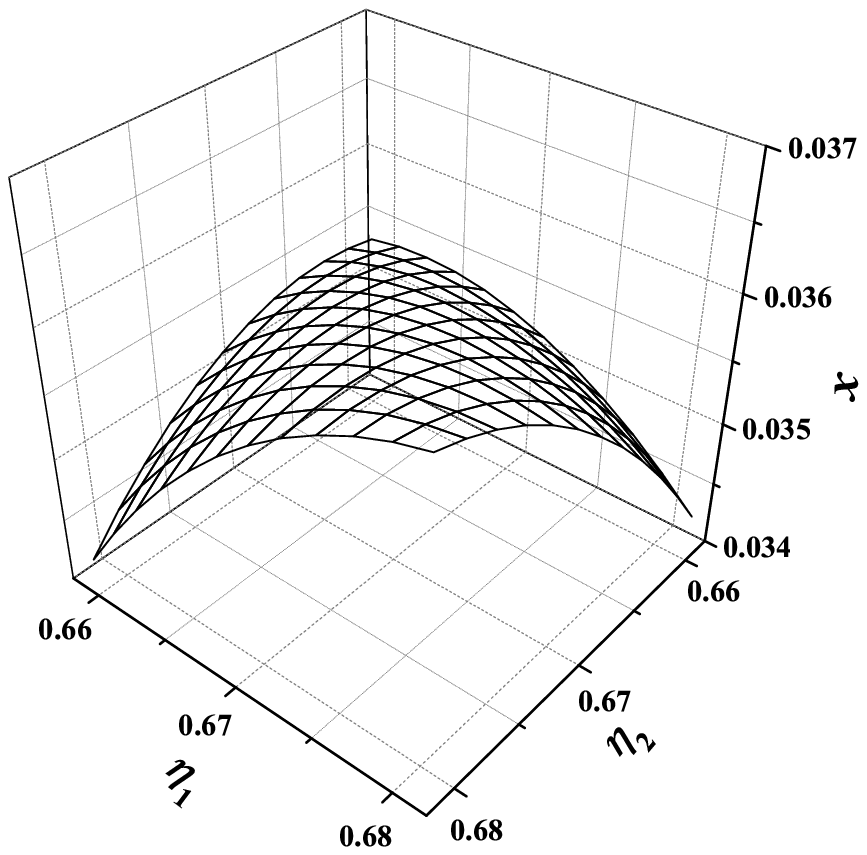}\hspace{1mm}
\includegraphics[width=.42\textwidth]{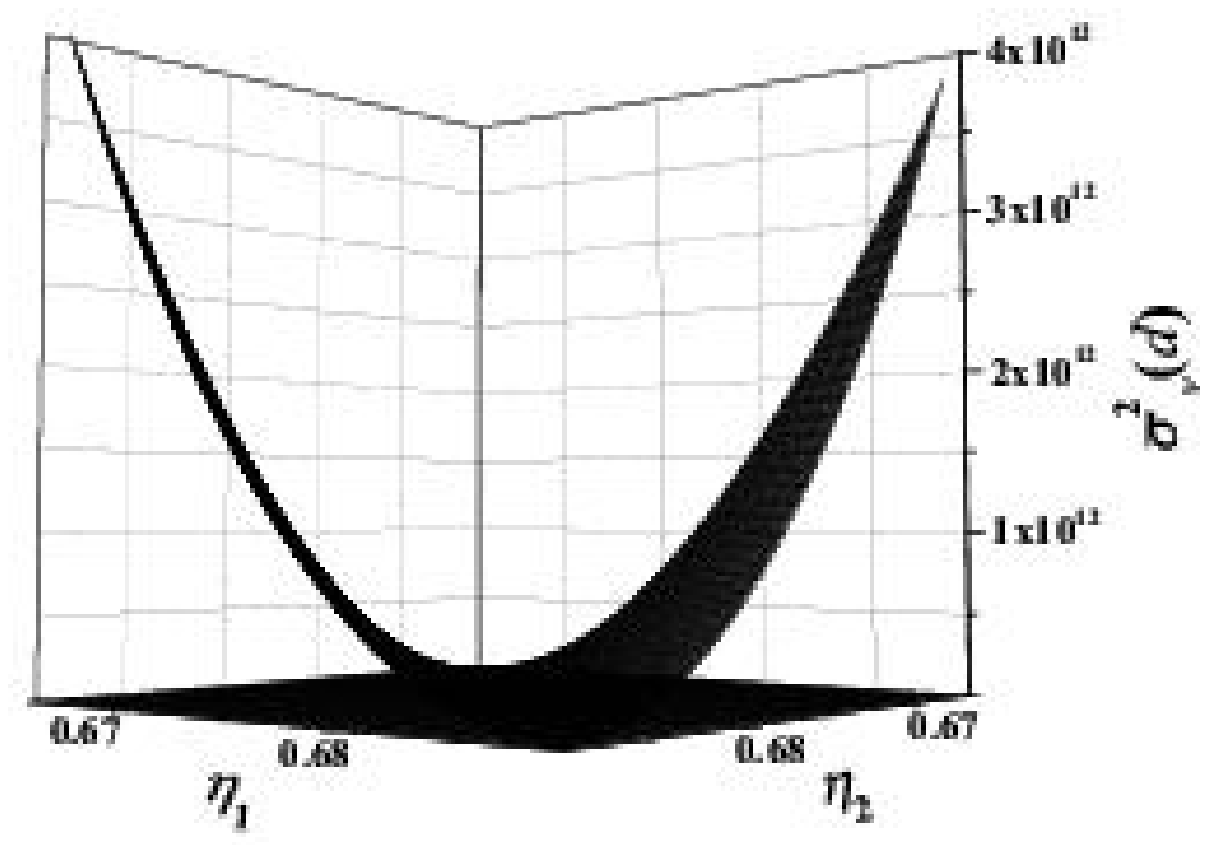}
\caption{Laser fluctuations in experiments with thermal light.
Left: the amount of laser fluctuations $x$ as a function 
of the quantum efficiencies $\eta_1$ and $\eta_2$. Right: values of 
the corrected variance $\overline{\sigma}^2_{\nu}(d)$ as a function 
of the quantum efficiencies $\eta_1$ and $\eta_2$; the
plane represents the shot-noise value.}
\label{f:xPCLASS}
\end{figure}

As concerning the thermal light experiments, by applying the same 
strategy, we find that the excess noise can be
written as
\begin{eqnarray}
\delta^2_{\nu} (n)&=& 2 x^2 N_{\nu}^2\ , \label{eq:DeltaNtherm}
\end{eqnarray}
which is again a function of the laser fluctuations $x$. Again we 
equate the value of the measured $\sigma^2_{\nu}(d)$ corrected for the added noise
$\delta^2_{\nu} (n)$  and study the dependence of $x$ on $\eta_1$
and $\eta_2$.
Figure~\ref{f:xPCLASS} displays the values of $x$ as a function of
$\eta_1$ and $\eta_2$ (left), and the corresponding values of the
corrected $\overline{\sigma}^2_{\nu}(d)$ (right). The horizontal
plane on the right represents the shot-noise level of the measure as
calculated from Eq.~(\ref{vara}). The highest values of the laser
fluctuations, which are found for $\eta_1\simeq \eta_2$, is $x\simeq
3.6\%$ at most. In contrast with the case of the TWB, from
Fig.~\ref{f:xPCLASS} we see that the values of
$\overline{\sigma}^2_{\nu}(d)$ are always above the horizontal plane
representing the shot-noise level of the measure.
\par
\begin{figure}[h]
\includegraphics[width=.42\textwidth]{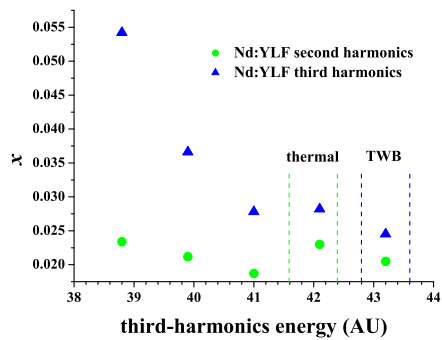}
\caption{Measured values of the laser fluctuations 
$x$ for second- and third-harmonics
outputs of the laser as a function of the third-harmonics energy in
arbitrary units. The vertical lines delimitate the energy ranges of
measurements performed on TWB an thermal light.} \label{f:xMIS}
\end{figure}

In order to check the plausibility of the calculated values of $x$, we
perform a stability measurements on the laser, by simultaneously
detecting the second- and third-harmonics outputs of the laser with
two photodiodes. In Fig.~\ref{f:xMIS} we plot the measured values of
$x$ as a function of the third-harmonics energy in arbitrary units.
The marked energy intervals in the plot indicate the operating range
of the measurements discussed above. The obtained values of $x$ are
in agreement with those calculated.
\section{Conclusion}\label{s:outro}
Establishing the existence of entanglement and discriminating
between classically and quantum correlated states in the
high-intensity continuous-variable regime is a challenging task
motivated by the need of characterizing the nature of the
correlated light and of understanding the real resources needed to
achieve the results in specific situations. We demonstrate that the 
characterization in
terms of correlation functions is not satisfactory, as it gives
similar results in both classical and quantum domain, whereas the
measurement of the probability distribution for the difference
photocurrent is in principle a good strategy. On the other hand, 
we demonstrate that in realistic high intensity conditions
such a strategy cannot be reliably adopted, due to the unavoidable
fluctuations of the laser source and slight imbalance of the 
detectors' quantum efficiencies. Indeed, by correcting the
experimental data for these sources of noise, the data analysis
leads to an agreement with the expected results.
\par
To achieve a more direct experimental demonstration we can follow 
two strategies. On one hand, we could work with identical quantum 
efficiencies, \textit{i.e.} at frequency
degeneracy, and use the same detection system on both 
parties of the correlated state. This could be done, for instance, by
substituting the p-i-n photodiodes with a CCD camera. 
On the other hand, 
one may lower the intensity of the field to be measured, 
to decrease the sensitivity to the excess noise due to the pumping laser.
Notice that, however, the possibility of lowering the intensity is limited
by the amplifying capability of the electronic chain that
manipulates the photodiode outputs. To overcome this limitation, 
one should switch to detectors with internal
gain, such as photomultiplier tubes and hybrid photodetectors, taking 
into account that these detectors shows a low quantum efficiency of
the photoelectric emission of the photocathodes which may 
compromises the overall visibility.
\par
In conclusion we have shown that difference photocurrent may be 
used, in principle, in order to reveal entanglement, while the simple
measurement of the correlation coefficient is not sufficient. 
Our experimental results indicate that 
joint photodetection may be useful to discriminate the
entanglement of twin-beam from correlations of thermal sources in
the mesoscopic regime.
\section*{Acknowledgments}
This work has been supported by MIUR (FIRB RBAU014CLC-002) and by
INFM (PRA-CLON). The Authors thanks F. Ferri for stimulating
discussion on the statistics of thermal light and E. Gevinti, P.
Rindi, E. Puddu and G. Zambra for technical support during the
measurements.

\end{document}